# HOW TO FIND METAL-RICH ASTEROIDS

Alan W. Harris and Line Drube

German Aerospace Center (DLR) Institute of Planetary Research, Rutherfordstrasse 2, 12489 Berlin, Germany

alan.harris@dlr.de

ABSTRACT

The metal content of asteroids is of great interest, not only for theories of their origins and the evolution of the Solar System but, in the case of near-Earth objects (NEOs), also for impact mitigation planning and endeavors in the field of planetary resources. However, since the reflection spectra of metallic asteroids are largely featureless, it is difficult to identify them and relatively few are known. We show how data from the WISE/NEOWISE thermal-infrared survey and similar surveys, fitted with a simple thermal model, can reveal objects likely to be metal rich. We provide a list of candidate metal-rich NEOs. Our results imply that future infrared surveys with the appropriate instrumentation could discover many more metal-rich asteroids, providing valuable data for assessment of the impact hazard and the potential of NEOs as reservoirs of vital materials for future interplanetary space activities and, eventually perhaps, for use on Earth.

1. INTRODUCTION

Asteroids assigned to the M taxonomic type may be largely Ni-Fe metallic fragments of the cores of differentiated bodies destroyed in collisions, although in general their mineralogical composition is certainly more complex (Rivkin, et al. 2000), and some may be relatively metal poor (Magri et al. 2007). Since the reflection spectra of M-type asteroids are largely featureless, it is relatively difficult to identify them; some 40 have been identified in the main asteroid belt on the basis of optical spectroscopy and visual albedos in the range $p_v \sim 0.1 - 0.3$ (Tholen 1984; Tholen & Barucci 1989). It is quite possible that metallic objects have been assigned to other taxonomic types with relatively featureless reflection spectra, such as C, E, or P. In early taxonomic systems the visual geometric albedo, $p_v$, was used as a classification parameter in addition to features in visual reflection spectra. However, later systems, such as the Small Main-belt Asteroid Spectroscopy Survey (SMASS) taxonomic system (Bus & Binzel 2002), extended into the infrared by DeMeo et al. (2009), do not use albedo as a classification parameter, therefore M-types cannot be distinguished from E and P types in these systems as all three types have similar featureless visual reflection spectra. Asteroids with such spectra are grouped together as various X types in the SMASS and DeMeo et al. systems, distinguished by subscript letters to signify the presence of subtle spectral features. Therefore the X class contains asteroids classed as E, M, and P by earlier schemes.

Furthermore, we note that composite objects consisting of silicates mixed with large amounts of metal almost certainly exist (e.g. see Shepard et al. 2010), and some S- and C-type asteroids have relatively high radar albedos (e.g. see Magri et al. 2007), therefore many relatively metal-rich objects may have taxonomic classes other than M.

There is abundant evidence, from telescope observations and meteorite samples, that the strength and porosity of NEOs is related to taxonomic type and mineral composition, with carbonaceous C-type asteroids expected to be more porous and less robust than silicaceous S-types or M types. Impact events also suggest that there is wide diversity in the structural robustness of impactors, with airbursts (e.g. Chelyabinsk, Tunguska) caused by objects that disrupt violently in the atmosphere, while similarly sized objects with a largely metallic composition have a relatively high density and strength and can reach the ground to form a crater (e.g. Barringer Crater, Arizona), thus causing much greater damage. Furthermore, there is currently considerable interest in the potential of NEOs as sources of raw materials, including various metals, for future activities in space and, eventually perhaps, for use on Earth (see, for example, Elvis 2012). NASA's Asteroid Redirect Mission concept (ARM; e.g. Mazanek et al. 2013) involves capturing a small asteroid and returning it to lunar orbit for in-situ investigations of relevance to planetary defense and resources exploitation.

For these reasons it is important to improve our knowledge of the mineralogical compositions of asteroids, and their potential as future sources of economically valuable raw materials. In this context we note that while the missions Hayabusa (Fujiwara et al. 2006) and Stardust (Brownlee et al. 2006) have returned material samples to Earth from the S-type NEO Itokawa and comet 81P/Wild 2, respectively, and further sample-return missions (e.g. Hayabusa-2, Saiki et al. 2013; OSIRIS-REx, Lauretta et al. 2012), are currently under development, no object thought to be metal rich is amongst their targets. Furthermore, due to their robustness and ability to survive passage through the atmosphere, metallic objects are disproportionately well represented in meteorite collections on Earth. Understanding the true number of metal-rich objects in the Solar System remains a longstanding problem.

## 2. THERMAL MODELING

The Near-Earth Asteroid Thermal Model (NEATM; Harris 1998), based on spherical geometry, offers a relatively straightforward means of deriving reasonably accurate diameters and albedos of asteroids and trans-Neptunian objects (in principle all atmosphereless bodies) from thermal-infrared data for objects with unknown physical characteristics. With near-Earth objects in mind, the NEATM was designed to extend the applicability of the thermal model concept described by Lebofsky et al. (1986) to objects with significant thermal inertia and/or rapid rotation. The model incorporates a fitting parameter, $\eta$, referred to for historical reasons as the beaming parameter, which allows the model surface temperature distribution to be adjusted to take account of the effects of thermal inertia, spin vector, and surface roughness, thereby giving a better fit of the model fluxes to the measurements. The temperature distribution over the illuminated hemisphere is given by $T(\theta, \varphi) = T_{SS} \cos^{1/4}\theta \cos^{1/4}\varphi$, where $\theta$ and $\varphi$ are latitude and longitude, the latter measured from the subsolar point, and $T_{SS} = [(1-A)S/(\eta\varepsilon\sigma)]^{1/4}$, from the condition for thermal equilibrium, where $A$ is the bolometric Bond albedo, $S$ the solar flux at the asteroid, $\varepsilon$ the emissivity, and $\sigma$ the Stefan-Boltzmann constant. The NEATM has zero thermal emission on the night side. A rough surface gives rise to a higher measured subsolar temperature than expected for a smooth surface, due to enhanced emission from surface elements that happen to be facing the Sun (the "beaming" effect).

The best-fit $\eta$ value is a measure of the departure of an asteroid's temperature distribution from that of an object with a smooth surface and zero thermal inertia, or zero spin, in thermal equilibrium with insolation (in which case $\eta = 1$). A number of investigators (e.g. Delbo' et al. 2003; Harris 2006; Wolters et al. 2005; Delbo' et al. 2007; Emery & Lim 2011, Lellouch et al.

2013) have used best-fit $\eta$ values from the NEATM as a guide to thermal inertia, defined as $\Gamma = (\kappa\rho c)^{0.5}$, where $\kappa$ is the thermal conductivity, $\rho$ the density, and $c$ the specific heat of the material. Gaffey (1989) used IRAS thermal-infrared band ratios as an indicator of metal abundance on main-belt asteroid surfaces. However, to date no systematic study has been carried out of the potential of $\eta$, as derived by NEATM model fitting, as a tracer of metal content in asteroids.

For random spin-vector orientations, best-fit values of $\eta$ can be taken as a rough proxy for the "thermal parameter", $\Theta = \Gamma\omega^{0.5}(\varepsilon\sigma T_{SS}^3)^{-1}$, where $\Gamma$ is the thermal inertia and $\omega$ the spin frequency (Spencer et al. 1989). In practice, best-fit values of $\eta$ range from about 0.5 for an object with a rough surface and very low thermal inertia, and/or very low spin rate, in which case $\Theta \sim 0$, to around 3.0 for a rapidly spinning object with very high thermal inertia, in which case $\Theta \gg 1$. An additional issue in the case of NEOs is the large solar phase angles at which many observations are made. For fixed, moderate thermal parameter, $\eta$ increases with solar phase angle (e.g. see Mainzer et al. 2011a, Wolters et al. 2008) due to a reduction in temperature in the subsolar region and thermal emission arising on the night side (although the increase in $\eta$ with phase angle is less pronounced in the case of very high values of the thermal parameter due to the thermal emission being spread around the asteroid). It should be borne in mind that the dependence of $\eta$ on solar phase angle can confuse attempts to interpret $\eta$ values in terms of $\Theta$ and thermal inertia at high phase angles. In contrast to $\Theta$, which is defined as a property of the body itself by setting the temperature equal to the subsolar equilibrium temperature (Spencer et al. 1989), $\eta$ also depends on the spin-axis orientation with respect to the solar direction and the degree of surface roughness. If the spin axis is oriented close to the solar direction, the surface temperature distribution and $\eta$ will be largely independent of thermal inertia and spin rate. However, in general the effects of thermal inertia and rotation cause $\eta$ to increase. Adding high thermal conductivity material, such as metal, to an object's surface would cause thermal inertia and $\eta$ to increase.

In order to accurately describe the surface temperature distribution of an asteroid, a more sophisticated thermophysical model is required. However, thermophysical analyses require information on asteroid shape and spin vector (e.g. see Spencer et al. 1989; Harris and Lagerros 2002), in addition to data taken at a number of different epochs, in order to provide a reliable value for thermal inertia; such models cannot be used for the purposes of initial survey-data analysis. The approach described here provides a simple means to identify which asteroids observed in a survey such as WISE are of interest for further study as potentially metal rich.

### 3. ANALSIS OF THERMAL-INFRARED SURVEY DATA

A survey of the sizes and albedos of over 100,000 asteroids using the NEATM has been carried out by the NASA WISE (Wide-field Infrared Survey Explorer) space telescope (Wright et al. 2010). WISE was launched to Earth-orbit in December 2009 carrying a 40-cm diameter telescope and infrared detectors. WISE surveyed the sky for 12 months and the objects observed included a total of at least 584 NEOs, of which more than 130 were new discoveries (Mainzer et al. 2011a). The fact that the cryogenic phase of the WISE mission measured asteroid thermal emission in up to 4 infrared bands, centered on 3.4, 4.6, 12, and 22 μm, allowed scientifically useful fitted $\eta$ values to be derived for many of the asteroids observed. NEATM fitting requires measurements of thermal emission in at least 2 bands, such as the 22 μm, 12 μm, and, for NEOs, 4.6 μm bands, of WISE (the 3.4 μm band allows the reflected solar radiation to be measured, which is important for albedo estimates). Therefore the WISE data enable the thermal inertia of asteroids to be investigated, at least in a statistical sense.

While a comprehensive quantitative analysis of the thermal-inertia information in the WISE data must await the final analysis and publication of the data, we have used the published WISE preliminary catalog to investigate the dependence of $\eta$, derived from fitting the NEATM to WISE main-belt asteroid flux data, on various parameters. The published WISE results for main-belt asteroids (Masiero et al. 2011) provide measurements of object diameter, geometric albedo $p_v$, near-infrared albedo $p_{IR}$ (i.e. the albedo in the 3.4 μm and 4.6 μm WISE bands, assumed to be constant over this spectral range), and $\eta$. Figure 1a shows $\eta$ plotted against $p_{IR}$ for some 3500 main-belt asteroids for which measurements of these parameters are available in the WISE published preliminary catalog (Masiero et al. 2011; objects for which default or assumed parameter values are given in the catalog have been excluded, as have objects with a fractional error in $\eta$ of more than 20%). The plot shows two main concentrations of points, broadly corresponding to the C-type asteroids ($p_{IR}$ < 0.1) and S-type asteroids ($p_{IR}$ > 0.3) (for discussions of the correspondence between $p_{IR}/p_v$ and taxonomic type see Mainzer et al. 2011b, 2012). Between these two concentrations of points there is a third, more diffuse, population. The red line superimposed on the plot represents the running mean of the 10 highest $\eta$ values in bins of 100 points. A remarkable feature in the maximum $\eta$ trend is the existence of a prominent peak which occurs at around $p_{IR}$ ~ 0.2 (the effect is much diluted if the overall mean is taken due to random spin vectors). A similar result was obtained plotting $\eta$ against visual geometric albedo, $p_v$, but the different taxonomic classes appear to separate better in $\eta/p_{IR}$ space.

In Fig. 1a it is evident that the peak in the mean-$\eta$ plot coincides with the region occupied by M-type asteroids. *We suggest that the higher values of $\eta$ in this region are due to large numbers of asteroids with enhanced surface thermal conductivity due to their metal content.* The fact that the peak persists after removal of the currently identified M types (either originally classified as such on the Tholen system or classified as X on the Bus-DeMeo system and having measured $p_v$ in the broad "M-type range" 0.075 - 0.3) implies that many more asteroids with high metal content are present in the main belt.

It should be remembered that the fitted $\eta$ values in the case of a rapidly spinning object with high thermal inertia may be low if the spin axis happens to be oriented near the solar direction. So low values of $\eta$ do not necessarily indicate low thermal inertia or spin rate. While spin vector, in addition to thermal inertia, contributes to $\eta$, we assume that the additional scatter in the $\eta$ values resulting from different spin vectors averages out in the 10-point means plotted in Fig. 1a so that meaningful bin-to-bin comparisons of mean $\eta$, as a proxy for thermal inertia, can be made.

An independent indication that $\eta$ values trace the metal content of asteroids is obtained from a comparison of $\eta$ with radar albedo. Metallic objects are strong radar reflectors. Figure 1b shows radar albedo plotted against WISE infrared albedo on the same infrared albedo scale as Fig. 1a. It is evident that the radar albedo data cluster into three distinct groups of points, which follow the clustering evident in Fig. 1a. Furthermore, the radar albedo distribution has a peak just below $p_{IR}$ = 0.2, close to the peak in the running-mean $\eta$ curve in Fig. 1a.

We have also investigated the $\eta$/radar albedo dependence in the case of NEOs. A search of the literature has revealed 12 objects with fitted $\eta$ values and measured radar albedos. Figure 2 shows a plot of $\eta$ versus radar albedo for NEOs, demonstrating a clear association of low/high values of $\eta$ with low/high values of radar albedo. In some cases values of $\eta$ quoted in the literature for the same object differ significantly, which is to be expected if observations are made at different epochs and therefore different aspect angles. An example appears to be 6178 (1986 DA), which has the highest radar albedo of any NEO measured to date (Ostro et al. 1991). Since we are interested here in the maximum value of $\eta$ in such cases, for the purposes of Fig. 2 we take the value of $\eta$ implied by fitting the NEATM to the measurements of Tedesco and Gradie (1987; see fig. 2a of Harris 1998) for 6178 (1986 DA), namely $\eta = 1.75 \pm 0.35$, giving diameter = 2.8 km and $p_v = 0.096$ for H = 15.9. In contrast, Mainzer et al. (2011a) derive the much lower value of $\eta = 0.87 \pm 0.1$ from WISE measurements, but corresponding values of diameter and $p_v$ (3.15 km and 0.078, respectively, from Pravec et al. 2012, for H = 15.9) are similar. The uncertainties in values of diameter and albedo derived from NEATM fitting are of the order of 15% and 30%, respectively (Mainzer et al. 2011a; Delbo' et al. 2003).

## 4. CANDIDATE METAL-RICH NEOs

As discussed in Sections 2 and 3, best-fit values of $\eta$ would be expected to vary significantly depending on the solar phase angle at which the observations were made, surface roughness, the spin vector, in addition to the amount of metal present on the surface. Most of the WISE observations of NEOs were made in the range $\alpha = 30º – 80º$, for which, according to Mainzer et al. (2011a), the median fitted $\eta$ value increases from around 1 to 1.5. In any case, values of $\eta$ well above 2 are strongly suggestive of unusually high thermal inertia.

Taking Fig. 1a for guidance, we have searched through the WISE NEO results to identify NEOs with $\eta$ values derived by NEATM fitting that may be metal rich. We filtered the WISE data to find objects with fitted $\eta > 2.0$ for any sighting, and $p_{IR}$ in the range 0.15 - 0.30 (cf. Fig. 1a). Table 1 lists the NEOs passing the filter. Ten of the 18 objects have published taxonomic classifications (e.g. see the EARN near-Earth asteroid physical properties database: http://earn.dlr.de), of which 7 are S or Q types and one is of type M, X, or D. While no published taxonomic type is available for 365071 (2009 AV), unpublished spectra from SMASS suggest an S-type classification, according to the Bus-DeMeo on-line taxonomy classification tool (http://smass.mit.edu/cgi-bin/busdemeoclass-cgi). Of the 18 objects 9 are potentially hazardous asteroids (PHAs). A similar result is obtained if the $p_v$ range $0.10 – 0.30$ is used in the filter instead of $p_{IR}$. Optimization of the filter will require further study; other parameters, such as $p_{IR}/p_v$ (see Mainzer et al. 2011a, b), may also be useful for filtering purposes. Objects identified in this way are potentially important targets for further optical, thermal-infrared and radar observations, and thermophysical modeling, in order to improve our knowledge of the factors governing NEO thermal and spectral properties, especially metal content.

**Table 1**

Candidate High Metal Content NEOs Passing the Filter $0.15 < p_{IR} \leq 0.3$; $\eta > 2.0$

| NEO | Tax. | PHA? | D (km) | $p_v$ | $\eta$ | $\eta_{err}$ | $p_{IR}$ |
|---|---|---|---|---|---|---|---|
| 138359 | | | 1.09 | 0.10 | 2.931 | 0.105 | 0.19 |
| 1865 | S | | 1.61 | 0.14 | 2.902 | 0.036 | 0.27 |
| 152931 | Q | | 1.65 | 0.24 | 2.884 | 0.138 | 0.24 |
| 152978 | S: | Y | 0.53 | 0.11 | 2.641 | 0.116 | 0.19 |
| 365071 | | Y | 0.87 | 0.15 | 2.559 | 0.117 | 0.28 |
| 3554 | X, M, D | | 3.05 | 0.09 | 2.411 | 0.072 | 0.19 |
| 215442 | | | 0.79 | 0.15 | 2.328 | 0.325 | 0.17 |
| 152558 | S | | 1.36 | 0.18 | 2.284 | 0.051 | 0.28 |
| 366774 | AS | Y | 0.86 | 0.20 | 2.284 | 0.059 | 0.29 |
| 250680 | | Y | 0.40 | 0.15 | 2.279 | 0.075 | 0.30 |
| 7822 | S | Y | 1.21 | 0.13 | 2.261 | 0.049 | 0.30 |
| 163243 | S, Q | Y | 1.68 | 0.17 | 2.191 | 0.061 | 0.23 |
| 263976 | L | Y | 0.79 | 0.13 | 2.165 | 0.043 | 0.18 |
| 142464 | | | 0.89 | 0.12 | 2.139 | 0.044 | 0.22 |
| 2002 NW16 | | | 0.85 | 0.16 | 2.118 | 0.066 | 0.26 |
| 103067 | S | Y | 1.28 | 0.25 | 2.114 | 0.056 | 0.29 |
| 363024 | | Y | 0.56 | 0.10 | 2.055 | 0.067 | 0.24 |
| 325102 | | | 0.36 | 0.12 | 2.048 | 0.063 | 0.18 |

Notes: Taxonomic classifications are taken from the EARN database (http://earn.dlr.de/). A colon following the taxonomic class signifies an uncertain classification. All WISE sightings in the catalog, including multiple sightings of the same object, have been considered for the purposes of this table; due to measurement errors or different observational circumstances one record in the WISE catalog may pass the filter while another for the same object may not. The uncertainties in values of diameter and albedo derived from NEATM fitting are of the order of 15% and 30%, respectively (Mainzer et al. 2011a; Delbo' et al. 2003). For details of the NEOWISE data see Mainzer et al. (2011a).

## 5. SUMMARY AND CONCLUSIONS

A threatening NEO containing a large amount of metal would presumably be relatively robust and massive, depending on its internal structure, factors that would require careful consideration by deflection-mission planners and/or those mandated to manage mitigation, e.g. evacuation, activities on the ground in advance of a possible impact. Moreover, the identification of NEOs with high metal content is an important task for recently-announced endeavors in the field of planetary resources. However, the identification of metallic asteroids by means of reflectance spectroscopy is difficult due to the lack of prominent characteristic features, thus the amount of metal in the NEO population is very uncertain. By comparing values of the NEATM fitting parameter, $\eta$, derived from WISE data with asteroid taxonomic classifications and radar data, we have shown that $\eta$ appears to be a useful indicator of asteroids containing large amounts of metal. We have identified 18 NEOs on the basis of their $\eta$ values and infrared albedos that are potentially metal rich. Objects identified in this way are important targets for further optical, thermal-infrared and radar observations, and thermophysical modeling, in order to improve our knowledge of the factors governing NEO thermal and spectral properties, especially metal content. The results presented here, based on the published WISE preliminary catalog, imply that next-generation surveys, provided they are equipped with sensors operating at multiple thermally-dominated infrared wavelengths, could provide a valuable indication as to which discoveries warrant further investigation regarding possible high metal content.

This publication makes use of data products from NEOWISE, which is a project of the Jet Propulsion Laboratory/California Institute of Technology, funded by the Planetary Science Division of the National Aeronautics and Space Administration. The research leading to these results has received funding from the European Union's Seventh Framework Programme (FP7/2007-2013) under grant agreement no. 282703 (the NEOShield Project). We thank the anonymous referee for her/his valuable comments that helped us to improve the presentation of the work.

FIGURE CAPTIONS

**Figure 1:**

**a.** WISE $\eta$ values versus infrared albedo for main-belt asteroids. Basic taxonomic types are shown as colored bullets. The C and S types are from the SMASS taxonomic system (Bus & Binzel 2002). The red curve is a plot of the mean of the highest 10 $\eta$ values in bins of 100 data points; the purple curve is the same after removal of all the currently identified M types from the dataset (either originally classified as such on the Tholen system or classified as X on the Bus-DeMeo system and having $p_v$ in the range 0.075 - 0.3). Objects with default values of $p_{IR}$ in the WISE dataset have been excluded, as have objects with fractional uncertainties in $\eta$ and $p_{IR}$ exceeding 20%. **b.** Radar albedo versus WISE near-infrared albedo, $p_{IR}$, for main-belt asteroids. The radar data are from Magri et al. (2007) and Shepard et al. (2010, 2012). The broad clustering into 3 groups seen in Fig. 1a is also evident here, despite the relatively small number of points, whereby the central group here corresponds to high radar albedo, and in Fig. 1a to a peak in $\eta$ and the location of the M types.

**Figure 2:**

Fitted $\eta$ values for NEOs versus radar albedo. The tendency for $\eta$ to correlate with radar albedo suggests that high $\eta$ values are indicative of high metal content. Note, however, that low values of $\eta$ do not necessarily indicate low thermal inertia due to random spin-axis orientations (see text). The NASA Planetary Data System (PDS) data set EAR-A-5-DDR-RADAR-V18.0 was used as a guide to available data but radar albedo values were taken from original sources, with additions from the compilation of L. Benner: http://echo.jpl.nasa.gov/~lance/asteroid_radar_properties/. Sources of $\eta$ values: Mainzer et al. (2011a), Harris et al. (1998); Harris & Davies (1999); Harris et al. (2005); Harris (1998). Data judged to be less reliable are plotted as open circles.

**Figure 1**

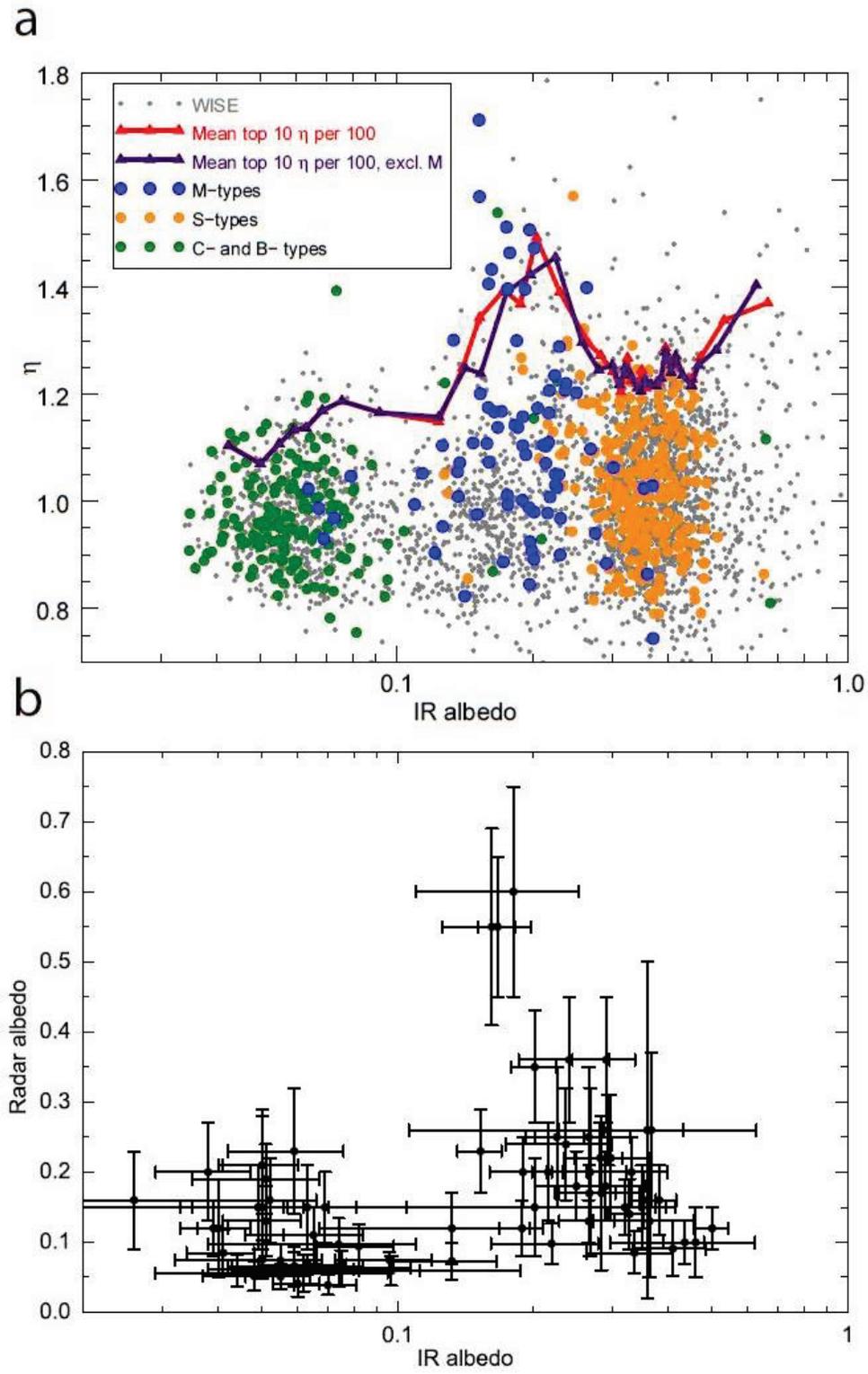

**Figure 2**

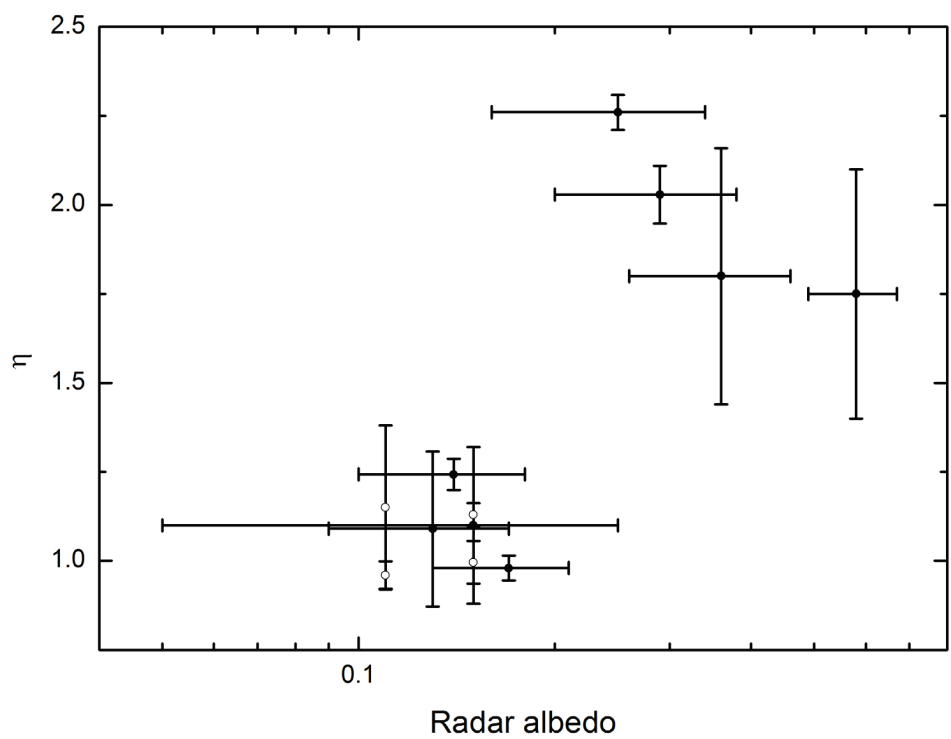